\documentclass{sig-alternate-2013}
\usepackage[normalem]{ulem}
\usepackage{amsmath}
\usepackage{graphicx}
\usepackage{amssymb}
\usepackage{epstopdf}
\usepackage{listings}
\usepackage{amsmath}
\usepackage{mathtools}
\usepackage{algpseudocode}
\usepackage{algorithm}
\usepackage{setspace}
\usepackage{tabularx}
\usepackage{url}
\usepackage{algorithm}
\usepackage{algpseudocode}

\newcommand{\combine}{\ensuremath{\mathsf{combine}}}
\newcommand{\remap}{\ensuremath{\mathsf{remap}}}

\newcommand{\ODS}{\ensuremath{\mathsf{ODS}}}
\newcommand{\PDL}{\ensuremath{\mathsf{PDL}}}

\newcommand{\AOS}{\ensuremath{\mathsf{AoS}}}
\newcommand{\forasync}{\ensuremath{\mathsf{forasync}}}

\newcommand{\SOA}{\ensuremath{\mathsf{SoA}}}

\newcommand{\Section}{\ensuremath{\mathsf{section}}}
\newcommand{\Sections}{\ensuremath{\mathsf{sections}}}
\newcommand{\ADHA}{\ensuremath{\mathsf{ADHA}}}
\algnotext{EndFor}
\algnotext{EndIf}
\algnotext{EndProcedure}
\newfont{\mycrnotice}{ptmr8t at 7pt}
\newfont{\myconfname}{ptmri8t at 7pt}
\let\crnotice\mycrnotice%
\let\confname\myconfname%

\permission{This is preprint of the paper submitted to PACT'14. Permission to make digital or hard copies of part or all of this work for personal or classroom use is granted without fee provided that copies are not made or distributed for profit or commercial advantage, and that copies bear this notice and the full citation on the first page. Copyrights for third-party components of this work must be honored. For all other uses, contact the owner/author(s). Copyright is held by the author/owner(s).}
\conferenceinfo{PACT'14,}{August 24--27, 2014, Edmonton, AB, Canada.}
\copyrightetc{ACM \the\acmcopyr}
\crdata{978-1-4503-2809-8/14/08. \\
http://dx.doi.org/10.1145/2628071.2628122}

\numberofauthors{4} %  in this sample file, there are a *total*
\author{
\alignauthor Deepak Majeti,\\ Kuldeep S. Meel\\
       \affaddr{Rice University}\\
       \email{\{deepak,~kuldeep\}@rice.edu}       
\alignauthor Rajkishore Barik\\
       \affaddr{Intel Labs}\\
       \email{rajkishore.barik@intel.com}
\and
\alignauthor Vivek Sarkar\\
       \affaddr{Rice University}\\
       \email{vsarkar@rice.edu}
}
\begin{document}

\title{ADHA: Automatic Data layout framework for Heterogeneous Architectures}
\maketitle
\begin{abstract}
%There is an increased awareness in recent work that 
Data layouts play a crucial role in determining the performance
of a given
application running on a given architecture. 
Existing parallel programming frameworks for both multicore and heterogeneous
systems leave the onus of selecting a data layout to the programmer. Therefore, shifting the burden of data layout selection to optimizing compilers can greatly enhance programmer productivity and application performance. 
In this work, we introduce {\ADHA}: a two-level
hierarchal formulation of the data layout problem for modern heterogeneous
architectures. 
We have created a reference implementation of {\ADHA} in the 
Heterogeneous Habanero-C (H2C) parallel programming system. {\ADHA}
shows significant performance benefits of up to 6.92$\times$ compared to  manually specified layouts for two
benchmark programs running on a CPU+GPU heterogeneous platform.
\end{abstract}
\vspace*{-3pt}
\category{D.3.4}{Software}{PROGRAMMING LANGUAGES}[Processors, Compilers]
\vspace*{-3pt}
\keywords{Compilers, Data Layout, Heterogeneous Architectures}
\section{Introduction}
In recent years, the end of Dennard scaling has brought about significant changes in the fundamental processor design. We are now entering an era of heterogeneous and specialized processors, and this trend is expected to continue in the future. One dominant heterogeneous architecture found in many systems today is a CPU+GPU system. The processor architecture, the memory hierarchy and cache structures are significantly different on the CPU and GPU sides. With such diverse characteristics, it not only hard to program these systems in a portable manner, but also quite challenging to optimize them.

One major factor which impacts performance is the data layout~\cite{dmajeti,sung}. The choice of a good data layout depends on several factors including target machine parallelism, memory hierarchy, data access patterns, and input size. An application program with multiple data-parallel kernels can map each kernel onto any of the heterogeneous processors. It is hard for the programmer to determine if a single layout is best for all the kernels or if a better choice is to select different data layouts for different kernels with data remapping operations performed in between kernels. Also, the programmer has to manually re-write the code for each data layout combination even to just evaluate the best data layout. The number of combinations increase exponentially with the number of kernels and the number of fields accessed by each kernel increase.

To overcome this limitation, we design {\ADHA}: a two-level compiler based automatic data layout framework and a
reference implementation of the same in the Heterogeneous Habanero-C~\cite{h2c} (H2C) programming system.
The lower level formulation deals with the data layout problem for a parallel code region, and 
provides a greedy algorithm that uses an {\em affinity graph} to obtain approximate solutions. 
 The higher level formulation targets data layouts for the entire program, 
for which we provide a graph-based shortest path algorithm that uses the data
layouts for the code regions computed in the lower level. In this work, we consider only \textit{AoS (Array of structure)} 
and \textit{SoA (Structure of Array)} layouts.
\section{Overall Framework}
We denote a data/task-parallel operation that is executed either on the CPU or on the GPU as a \Section. 
A H2C program may consist of several \forasync\ parallel loop constructs, each of which constitutes a \Section~of it's own.

Our automatic data layout framework, {\ADHA}, consists of two steps. The first step consists of a greedy strategy with the goal of determining the
 "optimal data-layout of a \Section" or simply {\ODS}.  The key idea is to construct an affinity 
graph for a H2C program where the "nodes" of the affinity graph represent the fields (and arrays) being accessed 
in the \Section~ and the "edges" represent the affinity between two fields. The affinity weight of an edge is computed
 based on the number of common occurrences between the two nodes involving the edge.
 Once the affinity graph is constructed, we employ a greedy clustering algorithm to cluster the fields. The cluster size is 
determined based on the underlying architecture (e.g., cache size and memory hierarchy). The result of the clustering
 algorithm is used to combine the fields in structures. 
 
 The next step involves finding the best data-layout for the entire H2C program (denoted as "program data-layout" or simply \PDL) 
from the computed best data layouts for each \Section. First,
 we construct a control flow graph for each \Section~(denoted as SCFG). %For now, we assume there are no branches involved in the SCFG. 
We then construct a rooted directed acyclic graph with the first \Section~as the root. Each pair of {\Sections} either share two edges, namely the \combine~edge and the \remap~edge, or none.
 The edge weight of the \combine~edge represents the loss in performance due to combining the two \Sections~involving
 the edge and assigning an intermediate data layout. The data layout of the combined \Sections~is obtained running the \ODS~pass on the merged \Section.
  
 The \remap~ edge weight is the cost of  remapping from the parent \Section~ to the child \Section~, which is computed  based 
on the number of common fields in the two \Sections. We then employ the shortest path algorithm to determine the best data layout 
for the entire program.
For the \PDL~pass, we provide a tuning profile of the execution times of both the CPU and GPU 
(since we do not perform the mapping automatically).
    The output of the \PDL~pass gives us the best data layout along with the mapping to the CPU or the GPU.

  Our implementation of {\ADHA} automatically compiles \forasync \\loops down to OpenCL 
with the corresponding data layout output by the \ODS~+ \PDL~ pass. {\ADHA} can be employed to efficiently run a
 H2C program on modern CPU+GPU platforms that support OpenCL.
\section{Experimental Evaluation}
We evaluated our implementation of {\ADHA} with two benchmarks arising from different domains as summarized in
Table~\ref{tab:benchmarks}. \textbf{\#S} denotes the number of \Sections~and \textbf{\#F} denotes the number of fields in the entire program. 
 \textbf{Medical} is from the CDSC benchmark suite~\cite{cdsc-research} and \textbf{K-Means} is from the Rodinia suite~\cite{rodinia1}.
\begin{table}[!h]
\vspace{-5pt}
\begin{center}
\scalebox{0.9}{
\begin{tabular}{|l|c|c|c|}%l|l|l|}}
\hline
    {\bf Description} & {\bf \#S} &{\bf \#F} &{\bf Input} \\\hline %{\bf LOC} & {\bf Input Size} & {\bf Default layout} \\ \hline 
     \texttt{Medical} Image Registration& 7&6 &256$\times$256$\times$256\\ 
    \texttt{K-Means} Clustering Algorithm & 2&32 & 8388608\\\hline
\end{tabular}
}
\end{center}
\vspace{-10pt}
\caption{Benchmarks Description}
\label{tab:benchmarks}
\vspace{-5pt}
\end{table}

The experiments were conducted on a Intel-X5660 CPU with 6 cores running at 2.8GHz and a
 NVIDIA Tesla-M2050 GPU with 8 SMs running at 575 MHz.
We use gcc version 4.4.6 with O2 optimization level. 
%to compile the generated programs.
For each of the benchmarks, we executed the OpenCL code with the original data layout and with the automatically generated layout from {\ADHA} 
on both the CPU and GPU.
\begin{figure}
\begin{center}
\includegraphics[height=1.8in,width=3in]{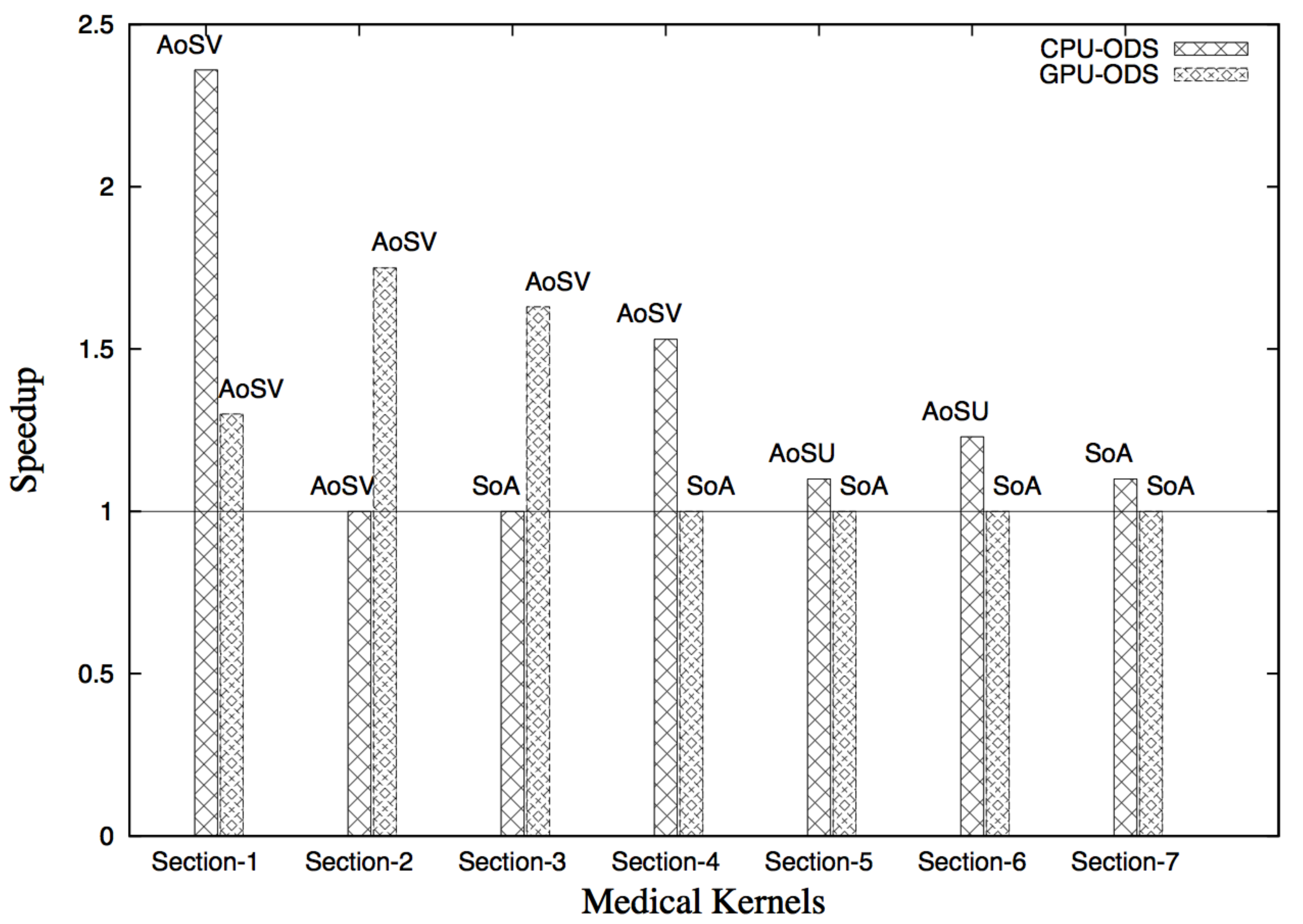}
\vspace*{-10pt}
\caption{Speedup of the Medical \Sections }
 \label{fig:medical}
 \end{center}
\vspace{-25pt} 
\end{figure}

\emph{Medical} image registration consists of 7 \Sections. Sections(1-3) have heavy control flow and do not vectorize while \Sections(4-7) are vectorizable. Table~\ref{medicaldatalayout} shows the fields accessed by all the \Sections~ and the different data layouts generated by the \ODS~ pass for each of the \Sections. For instance, $AoSU$ lays out fields $V1$,$V2$,$V3$,$S$,$T$ and $interpT$ as individual arrays using \SOA~layout and interleaves the fields $U1$,$U2$,$U3$ using an \AOS~layout. 
\begin{table}[h]
\vspace{-5pt}
\begin{center}
\scalebox{0.8}{
\begin{tabularx}{\columnwidth}{|c|X|}\hline
\textbf{Data Layout}&\textbf{Description}\\\hline
SoA&	V1,V2,V3,U1,U2,U3,S,T,interpT\\\hline
AoSU&	V1,V2,V3,\{U1,U2,U3\},S,T,interpT\\\hline
AoSV&	\{V1,V2,V3\},U1,U2,U3,S,T,interpT\\\hline
\end{tabularx}
}
\vspace{-5pt}
\caption{Medical Imaging ODS Data Layouts}
\label{medicaldatalayout}
\end{center}
\vspace{-10pt}
\end{table}
%\vspace*{-10pt}
Figure~\ref{fig:medical} shows the speedup from the data layout generated by the \ODS~pass compared to the original \SOA~layout by executing the seven \Sections~on CPU and GPU.
\ODS~pass performs better for most of the \Sections~on the CPU. 
On the GPU, the \SOA~layout performs the best for
 \Sections-(4-7) due to memory coalescing. The GPU benefits from cache locality for \Sections-(1-3).
   \begin{table}[!h]
    \begin{center}
\scalebox{0.8}{
    \begin{tabularx}{\columnwidth}{|c|c|X|}\hline
    \textbf{Platform}&\textbf{Layout Description}&\textbf{Speedup}\\\hline
    CPU-ODS&	32 Fields belong to \SOA&5.5\\\hline
    GPU-ODS&	4 \AOS~of size 8 each& 1\\\hline
    \end{tabularx}
    }
\vspace{-5pt}    
    \caption{KMeans Section-1 ODS and Speedup}
    \label{kmeansdatalayout}
    \end{center}
   \vspace*{-10pt}  
   \end{table}
 
   The benchmark \emph{K-Means} consists of two \Sections: the first \Section ~is a data parallel loop while the second \Section~ performs reduction on all the features. The second \Section~ is executed sequentially in the original implementation, owing to the difficulty of implementing reduction over varying number of variables using OpenCL. The original data layout for both these \Sections~is \SOA.
    The first \Section~CPU-ODS is an \AOS~ layout and the GPU-ODS is \SOA~layout and are shown along with their speedups in Table~\ref{kmeansdatalayout}. The \AOS~layout which is the output from the \ODS~ pass for the second \Section~improves its performance by a factor of \textbf{$8 \times$}.
    This is because the second \Section~suffers a lot of cache misses due to the \SOA~layout.

Table~\ref{pdl} provides the speedup obtained by the overall data-layout from the \PDL~pass. As stated before, the \PDL~pass gives the mapping onto CPU or GPU along with the data-layout. For the Medical benchmark, \Sections-(1-3) are mapped onto the CPU with $AOSV$ layout and then a remap operation is performed to $SOA$ layout with \Sections-(4-7) mapped to the GPU. 
\emph{K-Means} \Sections-(1-2) are both mapped onto the CPU with a combined \AOS ~layout.
    \begin{table}[!h]
     \begin{center}
\scalebox{0.8}{
     \begin{tabular}{|c|c|c|}\hline
     \textbf{Benchmark}&\textbf{\Section~Mapping SCFG}&\textbf{Speedup}\\\hline
     \textbf{Medical}&	CPU-ODS(1-3) \remap~GPU-ODS(4-7)& 1.34\\\hline
     \textbf{K-Means}&	CPU-ODS \combine~CPU-ODS& 6.92\\\hline
     \end{tabular}
     }
\vspace{-5pt}     
     \caption{PDL Speedup for Medical and K-Means}
     \label{pdl}
     \end{center}
     \vspace*{-15pt} 
     \end{table}
{
\bibliographystyle{IEEEtranS}
\bibliography{datalayout}

% Generated by IEEEtranS.bst, version: 1.13 (2008/09/30)
\begin{thebibliography}{1}
\providecommand{\url}[1]{#1}
\csname url@samestyle\endcsname
\providecommand{\newblock}{\relax}
\providecommand{\bibinfo}[2]{#2}
\providecommand{\BIBentrySTDinterwordspacing}{\spaceskip=0pt\relax}
\providecommand{\BIBentryALTinterwordstretchfactor}{4}
\providecommand{\BIBentryALTinterwordspacing}{\spaceskip=\fontdimen2\font plus
\BIBentryALTinterwordstretchfactor\fontdimen3\font minus
  \fontdimen4\font\relax}
\providecommand{\BIBforeignlanguage}[2]{{%
\expandafter\ifx\csname l@#1\endcsname\relax
\typeout{** WARNING: IEEEtranS.bst: No hyphenation pattern has been}%
\typeout{** loaded for the language `#1'. Using the pattern for}%
\typeout{** the default language instead.}%
\else
\language=\csname l@#1\endcsname
\fi
#2}}
\providecommand{\BIBdecl}{\relax}
\BIBdecl

\bibitem{cdsc-research}
\BIBentryALTinterwordspacing
``{CDSC Research Applications}.'' [Online]. Available:
  \url{http://www.cdsc.ucla.edu/research/}
\BIBentrySTDinterwordspacing

\bibitem{h2c}
\BIBentryALTinterwordspacing
``{Heterogeneous Habanero-C}.'' [Online]. Available:
  \url{http://habanero.rice.edu/Heterogeneous+Habanero-C}
\BIBentrySTDinterwordspacing

\bibitem{rodinia1}
Che \emph{et~al.}, ``Rodinia: A benchmark suite for heterogeneous computing,''
  ser. ISWC'09, Oct 2009.

\bibitem{dmajeti}
D.~Majeti \emph{et~al.}, ``{Compiler Driven Data Layout Transformation for
  Heterogeneous Platforms},'' in \emph{Proc. HeteroPar}, 2013.

\bibitem{sung}
I.-J. Sung \emph{et~al.}, ``Data layout transformation exploiting memory-level
  parallelism in structured grid many-core applications,'' in \emph{Proc. of
  PACT}, 2010.

\end{thebibliography}
}

\end{document}